\documentclass[twocolumn,amsmath,amssymb,prd,showpacs]{revtex4-2}

\usepackage{graphicx}
\usepackage{esvect}
\usepackage[mathscr]{euscript}
\usepackage{mathtools}
\usepackage{color}
\usepackage{float}
\usepackage{flushend}
\allowdisplaybreaks

\begin{document}

\def\al{\alpha}
\def\be{\beta}
\def\ga{\gamma}
\def\de{\delta}
\def\ep{\epsilon}
\def\ve{\varepsilon}
\def\ze{\zeta}
\def\et{\eta}
\def\th{\theta}
\def\vt{\vartheta}
\def\io{\iota}
\def\ka{\kappa}
\def\la{\lambda}\def\vpi{\varpi}
\def\rh{\rho}
\def\vr{\varrho}
\def\si{\sigma}
\def\vs{\varsigma}
\def\ta{\tau}
\def\up{\upsilon}
\def\ph{\phi}
\def\vp{\varphi}
\def\ch{\chi}
\def\ps{\psi}
\def\om{\omega}
\def\Ga{\Gamma}
\def\De{\Delta}
\def\Th{\Theta}
\def\La{\Lambda}
\def\Si{\Sigma}
\def\Up{\Upsilon}
\def\Ph{\Phi}
\def\Ps{\Psi}
\def\Om{\Omega}
\def\mn{{\mu\nu}}
\def\X{\hat X}
\def\Y{\hat Y}
\def\Z{\hat Z}
\def\E{\hat E}
\def\S{\hat S}
\def\U{\hat U}
\def\cA{{\cal A}}
\def\cl{{\cal L}}
\def\cE{{\cal E}}
\def\cI{{\cal I}}
\def\cN{{\cal N}}
\def\cS{{\cal S}}
\def\cO{{\cal O}}
\def\cB{{\cal B}}
\def\cC{{\cal C}}
\def\cF{{\cal F}}
\def\cG{{\cal G}}
\def\cM{{\cal M}}
\def\cP{{\cal P}}
\def\cR{{\cal R}}
\def\cV{{\cal V}}
\def\ap{\al^\prime}
	
\def\ol{\overline}
\def\ol#1{\overline{#1}}
\def\prt{\partial}
\def\pt{\phantom}
	
\def\ol{\overline}
\def\fr#1#2{{{#1} \over {#2}}}
\def\pt#1{\phantom{#1}}
\def\vev#1{\langle {#1}\rangle}
\def\ket#1{|{#1}\rangle}
\def\bra#1{\langle{#1}|}
\def\amp#1#2{\langle {#1}|{#2} \rangle}
\def\expect#1{\langle{#1}\rangle}
\def\norm#1{\left\|{#1}\right\|}
\def\abs#1{\left|{#1}\right|}
	
\def\half{{\textstyle{\frac{1}{2}}}}
\def\quar{{\textstyle{\frac{1}{4}}}}
\def\eigh{{\textstyle{\frac{1}{8}}}}
	
\def\lsim{\mathrel{\rlap{\lower4pt\hbox{\hskip1pt$\sim$}}
			\raise1pt\hbox{$<$}}}
\def\gsim{\mathrel{\rlap{\lower4pt\hbox{\hskip1pt$\sim$}}
			\raise1pt\hbox{$>$}}}
\def\sqr#1#2{{\vcenter{\vbox{\hrule height.#2pt
					\hbox{\vrule width.#2pt height#1pt \kern#1pt
						\vrule width.#2pt}
					\hrule height.#2pt}}}}
				
\def\Re{\hbox{Re}\,}
\def\Im{\hbox{Im}\,} 
\def\re{{\rm Re}~}
\def\im{{\rm Im}~}  
\def\ac#1{\left\{{#1}\right\}} 
\def\phihat{{\bf\hat\phi}} 
\def\rhohat{{\bf\hat\rho}} 
\def\sect#1{\vglue 0.6cm {\bf\ni {#1}} \vglue 0.4cm} 
\def\Lgr#1#2#3{L_{\scriptstyle #1}^{(\scriptstyle #2)}\left(#3\right)} 
\def\Gam#1{\Ga\left({\textstyle #1}\right)} 

\newcommand{\bM}{\begin{pmatrix}}
\newcommand{\eM}{\end{pmatrix}}
	
\newcommand{\beq}{\begin{equation}}
\newcommand{\eeq}{\end{equation}}
\newcommand{\bea}{\begin{eqnarray}}
\newcommand{\eea}{\end{eqnarray}}
\newcommand{\rf}[1]{(\ref{#1})}
\newcommand{\bit}{\begin{itemize}}
\newcommand{\eit}{\end{itemize}}
	
\def\nn{\nonumber}
	
\def\Psb{\ol\Ps{}}
\def\psb{\ol\ps{}}
	
\def\bsi{\ol\si}
\def\pb{\bar p}
\def\Omb{\widebar\Om}
	
\def\opr#1{\overrightarrow #1}
\def\opl#1{\overleftarrow #1}
\def\oplr#1{\overleftrightarrow #1}
	
\def\mbf#1{\boldsymbol #1}
	
\def\syjm#1#2{{}_{#1}Y_{#2}}
	
\def\O{\mathcal O}
\def\Q{\mathcal Q}
\def\S{\mathcal S}
\def\P{\mathcal P}
\def\V{\mathcal V}
\def\A{\mathcal A}
\def\T{\mathcal T}
\def\C{\mathcal C}
\def\G{\mathcal G}
\def\E{\mathcal E}
\def\B{\mathcal B}
\def\K{\mathcal K}
\def\H{\mathcal H}
\def\eom{E}
\def\eomz{{E_0}}
\def\BD{\Up}
\def\nv{N}
	
\def\sv{\Si}
\def\svvec{\mbf\Si}
\def\cEvec{\mbf\E}
\def\cBvec{\mbf\B}
\def\Avec{\mbf A}
\def\Evec{\mbf E}
\def\Bvec{\mbf B}
\def\pvec{\mbf p}
\def\pbvec{\mbf\pb}
\def\kvec{\mbf k}
\def\nvec{\mbf \nv}
\def\vvec{\mbf v}
\def\xvec{\mbf x}
\def\Pvec{\mbf P}
\def\gavec{\mbf\ga}
\def\thvec{\mbf\th}
\def\sivec{\mbf\si}
	
\def\del{\mbf\nabla}
	
\def\Jvec{\mbf J}
\def\Lvec{\mbf L}
\def\Svec{\mbf S}
\def\Kvec{\mbf K}
\def\Mvec{\mbf M}
\def\Tvec{\mbf T}
\def\bevec{\mbf\be}
\def\tavec{\mbf\ta}
\def\pchvec{\widetilde{\mbf p}}
\def\pch{\widetilde{p}}
	
\def\pmag{|\pvec|}
\def\komag{|{\mbf k_0}|}
\def\krmag#1{|{\mbf k_r}(#1)|}
	
\def\punit{\hat p}
\def\pbunit{\hat\pb}
\def\kunit{\hat k}
\def\xunit{\hat x}
\def\yunit{\hat y}
\def\zunit{\hat z}
\def\epunit{\hat\ep}
\def\thunit{\hat\th}
\def\phunit{\hat\ph}
	
\def\phat{\mbf\punit}
\def\pbhat{\mbf\pbunit}
\def\khat{\mbf\kunit}
\def\xhat{\mbf\xunit}
\def\yhat{\mbf\yunit}
\def\zhat{\mbf\zunit}
\def\ephat{\mbf\epunit}
\def\thhat{\mbf\thunit}
\def\phhat{\mbf\phunit}
	
\def\opr#1{\overrightarrow #1}
\def\opl#1{\overleftarrow #1}
\def\oplr#1{\overleftrightarrow #1}
	
\def\widebar{\overline}
\def\bigvev#1{\big\langle #1 \big\rangle}
\def\biggvev#1{\bigg\langle #1 \bigg\rangle}
\def\free{{\rm free}}
	
\def\cg#1#2{\langle #1 | #2 \rangle}
\def\bc#1#2{\left(\begin{smallmatrix} #1 \\ #2 \end{smallmatrix}\right)}
	
\def\Qhat{\widehat\Q}
\def\Shat{\widehat\S}
\def\Phat{\widehat\P}
\def\Vhat{\widehat\V}
\def\Ahat{\widehat\A}
\def\That{\widehat\T}
\def\Tdual{\widetilde{\widehat\T}\phantom{}}
\def\Chat{\widehat\C}
\def\Ghat{\widehat\G}
\def\Gahat{\widehat\Ga}
\def\Mhat{\widehat M}
	
\def\mhat{\widehat m}
\def\mfivehat{\widehat m_5}
\def\ahat{\widehat a}
\def\bhat{\widehat b}
\def\chat{\widehat c}
\def\dhat{\widehat d}
\def\ehat{\widehat e}
\def\fhat{\widehat f}
\def\ghat{\widehat g}
\def\Hhat{\widehat H}
\def\gdual{\widetilde{\widehat g}\phantom{}}
\def\Hdual{\widetilde{\widehat H}\phantom{}}
\def\gt{\widetilde g}
\def\Ht{\widetilde H}
	
\def\X{X}
\def\Y{Y}
\def\Z{Z}
\def\Xhat{\widehat\X}
\def\Yhat{\widehat\Y}
\def\Zhat{\widehat\Z}
	
\def\template#1#2#3#4{#1^{(#2)#4}_{#3}}
\def\acoef#1#2{\template{a}{#1}{#2}{}}
\def\bcoef#1#2{\template{b}{#1}{#2}{}}
\def\ccoef#1#2{\template{c}{#1}{#2}{}}
\def\dcoef#1#2{\template{d}{#1}{#2}{}}
\def\gzBcoef#1#2{\template{g}{#1}{#2}{(0B)}}
\def\goBcoef#1#2{\template{g}{#1}{#2}{(1B)}}
\def\goEcoef#1#2{\template{g}{#1}{#2}{(1E)}}
\def\HzBcoef#1#2{\template{H}{#1}{#2}{(0B)}}
\def\HoBcoef#1#2{\template{H}{#1}{#2}{(1B)}}
\def\HoEcoef#1#2{\template{H}{#1}{#2}{(1E)}}
\def\Template#1#2#3#4{(#1^{(#2)})^{(#3)}_{#4}}
\def\Edjm#1#2#3{\Template{\E}{#2}{#1}{#3}}
\def\Bdjm#1#2#3{\Template{\B}{#2}{#1}{#3}}
\def\nr{{\rm NR}}
\def\nrtemplate#1#2#3{#1^{\nr#3}_{#2}}
\def\anr#1{\nrtemplate{a}{#1}{}}
\def\bnr#1{\nrtemplate{b}{#1}{}}
\def\bbnr#1{\nrtemplate{b}{#1}{'}}
\def\cnr#1{\nrtemplate{c}{#1}{}}
\def\dnr#1{\nrtemplate{d}{#1}{}}
\def\ddnr#1{\nrtemplate{d}{#1}{'}}
\def\gzBnr#1{\nrtemplate{g}{#1}{(0B)}}
\def\goBnr#1{\nrtemplate{g}{#1}{(1B)}}
\def\goEnr#1{\nrtemplate{g}{#1}{(1E)}}
\def\HzBnr#1{\nrtemplate{H}{#1}{(0B)}}
\def\HoBnr#1{\nrtemplate{H}{#1}{(1B)}}
\def\HoEnr#1{\nrtemplate{H}{#1}{(1E)}}
\def\ur{{\rm UR}}
\def\urtemplate#1#2#3#4{#1^{\ur(#2)#4}_{#3}}
\def\aur#1#2{\urtemplate{a}{#1}{#2}{}}
\def\bur#1#2{\urtemplate{b}{#1}{#2}{}}
\def\cur#1#2{\urtemplate{c}{#1}{#2}{}}
\def\dur#1#2{\urtemplate{d}{#1}{#2}{}}
\def\gzBur#1#2{\urtemplate{g}{#1}{#2}{(0B)}}
\def\goBur#1#2{\urtemplate{g}{#1}{#2}{(1B)}}
\def\goEur#1#2{\urtemplate{g}{#1}{#2}{(1E)}}
\def\HzBur#1#2{\urtemplate{H}{#1}{#2}{(0B)}}
\def\HoBur#1#2{\urtemplate{H}{#1}{#2}{(1B)}}
\def\HoEur#1#2{\urtemplate{H}{#1}{#2}{(1E)}}
\def\ring#1{{\mathaccent'27 #1}}
\def\svfc{\ring\sv}
\def\fctemplate#1#2#3{\ring{#1}^{(#2)}_{#3}}
\def\afc#1#2{\fctemplate{a}{#1}{#2}}
\def\bfc#1#2{\fctemplate{b}{#1}{#2}}
\def\cfc#1#2{\fctemplate{c}{#1}{#2}}
\def\dfc#1#2{\fctemplate{d}{#1}{#2}}
\def\gfc#1#2{\fctemplate{g}{#1}{#2}}
\def\Hfc#1#2{\fctemplate{H}{#1}{#2}}
\def\nrfctemplate#1#2{\nrtemplate{\ring{#1}}{#2}{}}
\def\anrfc#1{\nrfctemplate{a}{#1}}
\def\bnrfc#1{\nrfctemplate{b}{#1}}
\def\cnrfc#1{\nrfctemplate{c}{#1}}
\def\dnrfc#1{\nrfctemplate{d}{#1}}
\def\gnrfc#1{\nrfctemplate{g}{#1}}
\def\Hnrfc#1{\nrfctemplate{H}{#1}}
\def\urfctemplate#1#2#3{\urtemplate{\ring{#1}}{#2}{#3}{}}
\def\aurfc#1#2{\urfctemplate{a}{#1}{#2}}
\def\burfc#1#2{\urfctemplate{b}{#1}{#2}}
\def\curfc#1#2{\urfctemplate{c}{#1}{#2}}
\def\durfc#1#2{\urfctemplate{d}{#1}{#2}}
\def\gurfc#1#2{\urfctemplate{g}{#1}{#2}}
\def\Hurfc#1#2{\urfctemplate{H}{#1}{#2}}
\def\Kcoef#1#2{\template{\K}{#1}{#2}{}}
\def\Knr#1{\nrtemplate{\K}{#1}{}}
\def\Kur#1#2{\urtemplate{\K}{#1}{#2}{}}
\def\Kfc#1#2{\fctemplate{\K}{#1}{#2}}
\def\Knrfc#1{\nrfctemplate{\K}{#1}{}}
\def\Kurfc#1{\urfctemplate{\K}{#1}{}}
	
\def\aclock#1#2{(\ring a_c)^{(#1)}_{#2}}
\def\cclock#1#2{(\ring c_c)^{(#1)}_{#2}}
\def\arod#1#2{(\ring a_r)^{(#1)}_{#2}}
\def\crod#1#2{(\ring c_r)^{(#1)}_{#2}}
	
\def\m{m_\ps}
	
%%%%%%%%%%%%%%%%%%%%%%%%%%%%%%%%%%%%%%%%%%%%%%%%%%%%%%%%%%%%%%%%%%%%%%%%%%%%%%%%%
	
\title{Quark-level CPT violation in neutral-meson systems}

\author{Ralf Lehnert and \'Agnes Roberts}

\affiliation{Indiana University Center for Spacetime Symmetries,
	Bloomington, Indiana 47405, USA}

\date{November 7, 2026}

\begin{abstract}
	This work investigates CPT and Lorentz breakdown
	in neutral-meson interferometry 
	within the nonminimal Standard-Model Extension.
	Employing methods of chiral perturbation theory for mass dimension $d=5$, 
	a connection between the quark-level and the meson effective pseudoscalar Lagrangian 
	is established.
	The resulting phenomenological signatures 
	in neutral-meson oscillations
	with particular focus 
	on the detector geometry 
	are explored. 
	This analysis 
	permits the interpretation of such CPT tests 
	within the $d=5$ quark sector of the Standard-Model Extension.
	Existing data
	are used to extract first constraints on two coefficients
	and to estimate achievable sensitivity gains over published bounds
	of up to 18 orders of magnitude.
\end{abstract} 

%\pacs{ {\color{red} PRELIMINARY DRAFT; NOT FOR DISTRIBUTION}}

\maketitle

%%%%%%%%%%%%%%%%%%%%%%%%%%%%%%%%%%%%%%%%%%%%%%%%%%%%%%%%%%%%%%%%%%%%%%%%%%%%%%%%%

\section{Introduction}
\label{Int}

Investigations of symmetries and their violation 
have been instrumental in the development of the Standard Model (SM) and remain of key significance 
for experimental and theoretical explorations 
of physics beyond the SM.
While most discrete spacetime symmetries 
have been established to be broken by the weak force,
there is no credible evidence for deviations
from the combined transformation of 
charge conjugation (C), 
parity inversion (P), 
and time reversal (T).
The theoretical basis of this symmetry 
is furnished by the celebrated CPT theorem~\cite{Schwinger,Bell,Lueders,Pauli,Jost,Greenberg:2003nv,Lehnert:2016zym}, 
which asserts 
that any realistic Lorentz-invariant quantum field theory 
must be CPT symmetric. 
This foundational result 
implies certain exact relations 
between any matter system 
and its antimatter twin.

On the other hand, 
the observable Universe displays
an imbalance between matter and antimatter. 
Despite containing the qualitative features 
for producing a matter--antimatter asymmetry,
the SM struggles to explain this
asymmetry quantitatively~\cite{CanettiKajantie,ParticleDataGroup:2024cfk}.
One avenue to address this question
involves minuscule departures from CPT symmetry~\cite{Bertolami96},
an effect that can be accommodated 
in various approaches to physics beyond the SM~\cite{strings,ng,foam,varying,other,finsler}.
This general idea has spawned 
the construction of the Standard-Model Extension (SME) test framework 
for the theoretical description of CPT violation
at the level of the SM and general relativity~\cite{KyCol,Ky04,KM,KL,Kostelecky:2018yfa}.
In the last two and a half decades,
these efforts have generated a surge of interest
in ultrahigh-precision CPT and Lorentz tests~\cite{tables},
including at particle accelerators~\cite{KyPotting95,NMPhKy98,NMPhKy99,NMPhKy01,KyRVKanomme,LHCb15,kaoncorrel,SchubBbar16,Babar08,BelleEPR07,AR17,BsKundu,KLOErecent,KLOE,BestD015,BaBar06,KTeV,LHCb16,FOCUS03,Belle12,Krzemien:2025sgt,ek19,Berger:2015yha,Kostelecky:2019fse,Lunghi:2020hxn,ZEUS:2022msi,Krzemien:2024eep,Belyaev:2024chj,Lunghi:2024ctv,other2},
with neutrinos~\cite{neutrinos},
and in astrophysics~\cite{astro}.

Neutral-meson interferometry has a long history
as a sensitive experimental tool in the study of discrete spacetime symmetries
including CPT invariance~\cite{ParticleDataGroup:2024cfk}.
Early theoretical descriptions of CPT breakdown 
in this context
were purely quantum mechanical~\cite{LeeWu,BigiSanda,classmod}.
In contrast,
the SME is an effective field theory,
so any CPT violation contained in the SME 
is subject to the constraints 
given by the ingredients of the CPT theorem.
In particular,
putative CPT violation in a realistic quantum field theory
is expected to come with Lorentz symmetry breaking~\cite{greenberg}.
For this reason,
departures from CPT invariance 
predicted in the SME 
are simultaneously also Lorentz violating
and therefore depend on the mesons' state of motion.
In the past,
this effect has provided the basis
for numerous theoretical and experimental meson-interferometry studies 
of CPT and Lorentz symmetry 
conducted within the minimal SME 
at mass dimensions $d=3,4$~\cite{KyPotting95,NMPhKy98,NMPhKy99,NMPhKy01,KyRVKanomme,LHCb15,kaoncorrel,SchubBbar16,Babar08,BelleEPR07,AR17,BsKundu,KLOErecent,KLOE,BestD015,BaBar06,KTeV,LHCb16,FOCUS03,Belle12,Krzemien:2025sgt}. 

However,
the full SME also includes effects at higher $d$ 
and thus contains additional CPT- and Lorentz-violating operators
affecting neutral-meson oscillations~\cite{KL}.
Recent efforts to extract their phenomenology 	were based on treating the composite mesons as pseudoscalar particles
disregarding their internal quark structure~\cite{ek19}. 
This approach is suitable 
to predict CPT-breaking effects
and can therefore be employed to place projected constraints 
using existing experimental data~\cite{ek19}.
But it is insufficient to 
interpret these constraints 
in terms of the microscopic quark SME coefficients 
for CPT and Lorentz violation.

Relating quark coefficients to observables 
is typically challenging
due to the involved strong-interaction physics. 
However,
various other studies 
including analyses involving
top-antitop pair production,
single-top production,
top and antitop mass reconstruction,
deep inelastic scattering, 
and the Drell-Yan process 
have been conducted within the SME
and have yielded experimental constraints on CPT and Lorentz violation in quarks~\cite{Berger:2015yha,Kostelecky:2019fse,Lunghi:2020hxn,ZEUS:2022msi,Belyaev:2024chj,Lunghi:2024ctv}.   
In this work, 
we employ a recently developed adaptation of chiral perturbation theory~\cite{ChPT1,ChPT2} 
that includes SME effects 
to relate quark- 
and meson-level CPT and Lorentz violation.
Our analysis thereby 
bridges the gap between nonminimal SME coefficients 
and the aforementioned phenomenological 
approach involving a pseudoscalar field. 
In addition, 
we apply our results 
to extend a previous phenomenological study  
of these effects
in neutral-meson interferometry 
by providing more detailed comments
on detector-geometry aspects 
of such measurements. 
Our work focuses 
on $a^{(5)}$-type operators in the SME's quark sector.
Previous studies of such operators~\cite{ZEUS:2022msi,Lunghi:2024ctv} 
have yielded constraints at the level of $10^{-5}-10^{-8}\,$GeV${}^{-1}$.
We conservatively estimate 
that dedicated analyses of existing meson-interferometry data
based on our results 
probe $a^{(5)}$ coefficients 
with sensitivities of $10^{-15}-10^{-25}\,$GeV${}^{-1}$.

%Our work focuses 
%on mass dimension $d=5$ CPT and Lorentz breakdown 
%and highlights analogies 
%to the well established $d=3$ minimal-SME case.

The structure of this work is as follows.
Section~\ref{Basics} places our study in the context
of existing results 
and formulates the question 
we aim to address. 
In Sec.~\ref{ChPT},
we discuss 
how to embed the generic $a^{(5)}$-type $d=5$ fermion coefficient 
for CPT and Lorentz violation
into the quark sector of the SME. 
We then study the leading-order ramifications 
of this $a^{(5)}$ SME quark sector 
for meson properties 
using adapted chiral perturbation theory 
and match the result to the pseudoscalar-field approach.
Phenomenological implications of these $a^{(5)}$ meson properties are discussed in Sec.~\ref{nonmin} 
with projected sensitivities of KLOE in Sec.~\ref{EstimKLOE}. 
A brief summary is contained in Sec.~\ref{Conclusion}. 
Throughout,
we work in natural units $c=\hbar=k_B=1$.

%%%%%%%%%%%%%%%%%%%%%%%%%%%%%%%%%%%%%%%%%%%%%%%%%%%%%%%%%%%%%%%%%%%%%%%%%%%%%%%%%

\section{Basics}
\label{Basics}

When studying the neutral-meson system, 
a practical start is to look at the description of meson oscillations 
via a Schrödinger-like equation 
and the corresponding effective Hamiltonian in the Weisskopf-Wigner approximation 
as it is traditionally studied 
within a quantum mechanical framework~\cite{LeeWu,BigiSanda,classmod}:
\beq
i\prt_t \Ps = \La \Ps ,
\label{schreq}
\eeq
where $\Ps(t)$ is the two-component wave function 
describing both the meson and the antimeson  
\beq
\Ps(t)= 
\begin{pmatrix}\al(t)\\ \be(t)\end{pmatrix},
\label{wavef}
\eeq 
while $\Lambda$ denotes the $2 \times 2$ effective Hamiltonian
governing the evolution of the system
in meson-antimeson state space 
in the system's rest frame.
Paralleling earlier studies of meson oscillations in the SME~\cite{NMPhKy01}, 
we adopt the parametrization
\beq
\La = 
\half \De\la
\left( \begin{array}{cc}
	U + \xi 
	& 
	\; VW^{-1} \\ 
	& \\		
	VW \; 
	& 
	U - \xi 
\end{array}
\right).
\label{uvwx}
\eeq
The eigenvalues $\lambda_a$ and $\lambda_b$ of this Hamiltonian 
determine the masses and decay rates of the two physical states.
Their difference is denoted by $\Delta \lambda$
and consequently given 
by the complex combination of the mass 
and decay-rate differences: 
$\Delta \lambda = -\De m-(i \De \ga)/2$.
In the above parametrization, 
$W$ and $\xi$ provide the complex, 
free parameters to describe discrete-symmetry violation in a phase-independent way~\cite{NMPhKy01}.
The quantities $U$ and $V$ can be related 
to $W$, $\xi$, $\lambda_a$, and $\lambda_b$. 
In particular, indirect CPT violation is governed 
by finite values of $\xi$.
As usual,
$\xi$ is tied to the diagonal-element difference $\Delta\Lambda=\frac{1}{2}(\Lambda_{11}-\Lambda_{22})$ 
of the Hamiltonian. In a first-order approximation in the coupling constant to the CPT-violating background \cite{KyPotting95,NMPhKy98,NMPhKy01}
\beq
\xi=\frac{\Delta\La}{\Delta\la}\,.
\label{xieq}
\eeq
Departures from discrete symmetries other than CPT
are encoded in different model parameters.
For example, the modulus $w$ of $W \equiv w \exp (i\om)$ 
governs T violation with $w = 1$ 
if and only if T is preserved.

We remark in passing 
that the parametrization~(\ref{uvwx}) 
is implicitly linearized 
based on the assumption 
that terms of higher order in $\xi$ 
are subdominant. 
Note also 
that even powers of $\xi$ would be CPT symmetric 
and thus fall outside our present focus.
However, 
${\cal O}(\xi^2)$ contributions are known 
to have interesting experimental implications
for correlated-meson measurements~\cite{AR17}.

Within this quantum mechanical approach,
a natural assumption has been 
that the entries of $\Lambda$ are constant 
for a given neutral-meson system. 
However,
$\Lambda$ provides merely an effective description of oscillations 
in such systems
and must therefore be understood 
as embedded into the underlying SM structure.
As such,
$\Lambda$ must for example remain compatible 
with both the basic tenets of quantum field theory 
and the composite nature of meson states at the quark level.
In particular,
the values of parameters in the Hamiltonian~(\ref{uvwx}) 
should arise from the underlying SM physics. 
Indeed, 
efforts along these lines have a long history 
and can readily be identified in the literature~\cite{Gaillard:1974hs,Buchalla:1995vs,Aoki:2016frl}.
In particular, 
a spacetime-constant $\Lambda$ 
is compatible with the SM
and captures the corresponding meson-oscillation physics well. 

However, 
the above reasoning 
meets an immediate obstacle
for the CPT-violating parameter $\xi$: 
the CPT theorem states 
that under mild technical assumptions 
Lorentz-invariant quantum field theories,
like the SM,
are CPT symmetric 
and therefore require $\xi=0$.  
A crucial question then becomes
how to extend the SM 
to incorporate perturbative CPT violation.
The enormous success 
of effective field theory (EFT)
for the description of a broad range of physical systems,
both with and without Lorentz symmetry, 
has spawned the paradigm 
that corrections to the SM 
can also be described within EFT.

This suggests amending the SM Lagrangian ${\cal L}_{\rm SM}$
by a correction $\delta{\cal L}$ 
incorporating CPT-odd effects,
${\cal L}_{\rm SME}={\cal L}_{\rm SM}+\delta{\cal L}$, 
to arrive at a broader EFT 
governed by ${\cal L}_{\rm SME}$.
It can then be argued 
that realistic choices of $\delta{\cal L}$ violate Lorentz invariance
to evade the CPT theorem~\cite{greenberg}.
Thus, 
CPT-violating effects such as those captured by $\xi$ 
are also associated with a loss of symmetry under rotations and boosts, 
so that $\xi$ must in general be orientation and velocity dependent 
for compatibility with an underlying EFT \cite{NMPhKy98}. 
%A reasoning along these lines 
%has been instrumental in the construction 
%of the aforementioned SME framework. 

The aforementioned SME is an EFT 
that classifies CPT- and Lorentz-violating corrections $\delta{\cal L}$ 
to the SM Lagrangian coupled to general relativity 
that remain compatible with other fundamental principles in physics.
The SME has thus provided a natural starting point 
for CPT-symmetry studies in neutral-meson systems.
For example,
the SME's quark Lagrangian 
contains a mass dimension $d=3$  contribution
given by
\begin{equation}
	\label{amu}
	{\cal L}\supset -\bar{q}\,a_q^{(3)\mu}\gamma_{\mu} q\,,
\end{equation}
where $q$ is a quark field, 
$\gamma_{\mu}$ denote the usual Dirac gamma matrices, 
and $a_q^{(3)\mu}$ represents a set of four constant SME coefficients
parametrizing the relevant type of Lorentz and CPT violation 
for the quark $q$. 

This EFT term 
is known to generate a nonzero contribution to the $\xi$ parameter 
of the form~\cite{NMPhKy98,KyPotting95}
\begin{equation}
	\label{xi_amu}
	\xi_P =\frac{\big(r_{q_{2}}a_{q_{2}}^{(3)\mu}-r_{q_{1}}a_{q_{1}}^{(3)\mu}\big)\zeta_{\mu}}{\De\la},
\end{equation}
with $P=K^0,D^0,B_d^0,B_s^0$ labeling the meson system in question,
$a_{q2}^{(3)\mu}$, $a_{q1}^{(3)\mu}$ the SME coefficients of the two valence quarks of $P$ 
expressed in a given frame,
and $\zeta^{\mu}$ the 4-velocity of the system 
with respect to that frame.
This result relies on a valence-quark approximation,
in which undetermined coefficients $r_{q_i}$ 
have been introduced to capture strong-interaction effects, 
such as those arising from sea quarks and gluons. 
In Sec.~\ref{ChPT} below,
we present the analogous analysis for the SME's $a^{(5)\mu\alpha\beta}_q$ coefficients,
but with strong-interaction effects incorporated instead 
via chiral perturbation theory.
	
The $d=3$ quark operator~(\ref{amu})
represents only a single type of CPT violation 
compatible with EFT.
More general EFT effects 
at mass dimensions $d>3$ 
provide more comprehensive probes of CPT symmetry, 
and their investigation was initiated in~\cite{ek19}.
That study treated a given neutral-meson system 
as a free scalar quantum field $\phi$ 
and parametrized all Lorentz- and CPT-violating operators 
of arbitrary mass dimension in this context:
\begin{equation}
\label{all_d_CPTV}
{\cal L}_{\rm CPT}\supset-\half\phi^\dagger(\widehat{k}_a)^\mu\partial_\mu\phi+\textrm{h.c.}\,,
\end{equation}
where
\begin{equation}
\label{k_a_def}
(\widehat{k}_a)^\mu\equiv
(k_a^{(3)})^\mu+
\sum_{d\ge 5}(k_a^{(d)})^{\mu\alpha_1\alpha_2\ldots\alpha_{d-3}}
\prod_{j=1}^{d-3}(i\partial_{\alpha_j})\,.
\end{equation}
Here, 
the coefficients $(k_a^{(d)})^{\mu\alpha_1\alpha_2\ldots\alpha_{d-3}}$ 
are taken as real, spacetime constant, traceless, and symmetric in all Lorentz indices. 
It follows that a given $(k_a^{(d)})^{\mu\alpha_1\alpha_2\ldots\alpha_{d-3}}$
contains $(d-1)^2$ independent components, 
and that it is measured in $\textrm{GeV}^{4-d}$.
At each mass dimension $d$, 
the phenomenological $\xi$ parameter 
receives a contribution~\cite{ek19}
\begin{equation}
\label{xi_d}
\xi^{(d)}=\dfrac{m^{d-3}}{\Delta\lambda}\zeta_{\alpha_1}\zeta_{\alpha_2}\ldots\zeta_{\alpha_{d-2}}(k_a^{(d)})^{\alpha_1\alpha_2\ldots\alpha_{d-2}}\,,
\end{equation}
where the components of $(k_a^{(d)})$ are specified in a given frame,
\beq\label{beta_expr}
\zeta^\mu=(\zeta^0,\vec{\zeta})=\gamma(1,\vec{\beta})
\eeq
is the meson's 4-velocity 
with respect to that frame,
and $\vec{\beta}$ denotes the corresponding 3-velocity.

This approach again exhibits 
the orientation and velocity dependence of CPT violation,
as expected,
and it permits the phenomenological study of CPT symmetry 
in neutral-meson oscillations 
compatible with realistic EFT. 
In fact, 
some experimental constraints on $(k_a^{(5)})$ 
are already available~\cite{ek19}. 
However,
the question as to how the $(k_a^{(d)})$
are related to the SME's more elementary quark and gluon coefficients 
remains largely unexplored. 
Comparison to the analogous $d=3$ SME result~(\ref{xi_amu}) reveals 
that %~\cite{ek19}%
\begin{equation}\label{k3expr}
(k_a^{(3)})^\mu \simeq 2(r_{q_{2}}a_{q_{2}}^{\mu}-r_{q_{1}}a_{q_{1}}^{\mu})\,,
\end{equation}
but the results for other $d$ are still unknown. 
In this work,
we explore the $d=5$ case 
and, in particular how $(k_a^{(5)})^{\mu\nu\rho}$
can arise from SME physics 
and what further phenomenological signatures it holds.
	
%%%%%%%%%%%%%%%%%%%%%%%%%%%%%%%%%%%%%%%%%%%%%%%%%%%%%%%%%%%%%%%%%%%%%%%%%%%%%%%%%	
	
\section{CPT violation at $\boldsymbol{d=5}$ in the SME}
%{\color{green}\section{Meson- and quark-level CPT violation at $d=5$ in the SME}}
\label{ChPT}

This section considers the question 
of how Lorentz- and CPT-breaking SME coefficients, 
which are assigned to elementary constituents of the SM such as quarks,
can generate an effective $(k_a^{(5)})^{\mu\nu\rho}$ contribution 
for a neutral-meson system. 
As we will be working at leading order in Lorentz and CPT violation,
$(k_a^{(5)})$ can only arise 
from traceless and nonaxial SME coefficients 
with the same number of spacetime indices 
and with associated discrete-symmetry properties 
matching those of the $(k_a^{(5)})$ operator 
in the scalar Lagrangian~\rf{all_d_CPTV}.

%$(k_a^{(5)})$ can only arise from traceless and non-axial SME coefficients 
%with the same number of spacetime indices 
%and from SME contributions with discrete-symmetry properties 
%matching those of the $(k_a^{(5)})$ operator in the scalar Lagrangian~\rf{all_d_CPTV}.

\subsection{Light-quark $\boldsymbol{d=5}$ Lagrangian}

Promising candidates with these properties  
are the SME's $a$-type fermion coefficients. 
These have already been identified for the minimal SME 
as the quark coefficient
most relevant for neutral-meson oscillations~\cite{NMPhKy98}. 
At $d=5$,
they generically enter the Lagrangian as follows~\cite{Kostelecky:2018yfa}:
\beq
\label{a5def}
{\cal L}\supset-\tfrac{1}{2}a^{(5)}{}^{\mu\alpha\beta}\bar{\psi}\gamma_{\mu}iD_{(\alpha}iD_{\beta)}\psi+\textrm{H.c.}
\eeq
Here,
$\psi$ denotes a fermion field, 
$\gamma_\mu$ the usual gamma matrices, 
and $D_\alpha$ a gauge-covariant derivative. 
The parentheses enclosing the two spacetime indices 
represent symmetrization with a factor of one half, 
and H.c.~abbreviates the Hermitian conjugate of the preceding term.

This contribution indeed exhibits the same behavior under C, P, and T transformations 
as the scalar-field $k_a^{(5)}$ correction,
the contraction of any two of its Lorentz indices may be set to zero, 
and it is also of mass dimension $d=5$.
However,
it is only symmetric in its last two indices, 
and so we expect that only 
\beq
\label{a5symdef}
a_s^{(5)}{}^{\mu\nu\rho}\equiv\tfrac{1}{3}\big(a^{(5)}{}^{\mu\nu\rho}+a^{(5)}{}^{\nu\rho\mu}+a^{(5)}{}^{\rho\mu\nu}\big)\;,
\eeq
which represents its fully symmetric piece,
will contribute to meson effective Lagrangians. 
This feature holds more generally 
and can be established using fermion-field redefinitions~\cite{Ding:2016lwt,Kostelecky:2019fse}. 

\begin{table}
	\begin{center}
		\begin{tabular}{cccc}
			\hline\hline
			Coefficient & \phantom{AA}{\it C}\phantom{AA} & \phantom{AA}{\it P}\phantom{AA} & \phantom{AA}{\it T}\phantom{AA}\\
			\hline
			$\vphantom{\vec{\vec{A}}}(k_a^{(5)})^{000}$, $a^{(5)}{}^{000}$ & $-$ & $+$ & $+$\\
			$(k_a^{(5)})^{00j}$, $a^{(5)}{}^{00j}$ & $-$ & $-$ & $-$\\
			$(k_a^{(5)})^{0jk}$, $a^{(5)}{}^{0jk}$ & $-$ & $+$ & $+$\\
			$(k_a^{(5)})^{jkl}$, $a^{(5)}{}^{jkl}$ & $-$ & $-$ & $-$\\
			\hline\hline
		\end{tabular}
		\caption{Lorentz-violating coefficients and associated {\it C}, {\it P}, and {\it T} parities.
			Note that the $k_a^{(5)}$ and $a^{(5)}$ contributions to EFT Lagrangians exhibit identical behavior under {\it C}, {\it P}, and {\it T}.
			Note also that all coefficients involve CPT breakdown.}
		\label{CPTproperties}
	\end{center}
\end{table}

In a first step,
we need to embed $a^{(5)}$-type coefficients into the gauge structure of the SM. 
Fermionic degrees of freedom directly enter meson physics 
via their quark substructure.
Other fermions may only have indirect effects,
e.g., 
as virtual particles,
and we disregard their contributions here.
For our present purposes,
we are thus led to include $a^{(5)}$ solely into the SM's quark sector. 

We assume that SU(3)$\times$SU(2)${}_L\times$U(1) gauge symmetry is maintained.
To display this explicitly,
we express these contributions in terms of 
the usual left-handed weak doublet $Q_A$ 
and the right-handed weak singlets $U_A$ and $D_A$,
where the subscripts $A,B\in\{1,2,3\}$ are the ordinary family labels: 
\begin{align}
	\label{quarkSME}
	\delta{\cal L}^{a^{(5)}}_{\rm qSME}=
	& {}-\tfrac{1}{2}(a_Q^{(5)})_{AB}^{\mu\alpha\beta}\bar{Q}_A\gamma_{\mu}iD_{(\alpha}iD_{\beta)}Q_B+\textrm{H.c.}\nonumber\\
	& {}-\tfrac{1}{2}(a_U^{(5)})_{AB}^{\mu\alpha\beta}\bar{U}_A\gamma_{\mu}iD_{(\alpha}iD_{\beta)}U_B+\textrm{H.c.}\nonumber\\
	& {}-\tfrac{1}{2}(a_D^{(5)})_{AB}^{\mu\alpha\beta}\bar{D}_A\gamma_{\mu}iD_{(\alpha}iD_{\beta)}D_B+\textrm{H.c.}
\end{align}
Without loss of generality,
we may choose the flavor matrices $(a_Q^{(5)})_{AB}$, $(a_U^{(5)})_{AB}$, and $(a_D^{(5)})_{AB}$ 
in $\delta{\cal L}^{a^{(5)}}_{\rm qSME}$ 
to be Hermitian. 
The $D_\alpha$ with Greek indices 
refer here to SU(3)$\times$SU(2)${}_L\times$U(1)-covariant derivatives.

The quark-level structure of mesons is primarily determined by the strong interaction, 
so we may disregard electroweak physics 
when studying leading-order $a^{(5)}$ effects. 
To determine these effects, 
we will employ chiral perturbation theory ($\chi$PT)~\cite{ChPT}.
As this method provides the best results for light quarks, 
we only retain the $u$, $d$, and $s$ degrees of freedom 
in $\delta{\cal L}^{a^{(5)}}_{\rm qSME}$ in what follows. 
Judiciously dropping surface terms, 
this yields
\begin{align}
	\label{udsLR}
	\delta{\cal L}^{a^{(5)}}_{\rm qSME}=
	&{}-\bar{\Psi}_L(a_L)^{\mu\alpha\beta}\gamma_{\mu}iD_{(\alpha}iD_{\beta)}\Psi_L\nonumber\\
	&{}-\bar{\Psi}_R(a_R)^{\mu\alpha\beta}\gamma_{\mu}iD_{(\alpha}iD_{\beta)}\Psi_R\,,
\end{align}
where we have collected the quark fields into left- and right-handed column vectors
$\Psi_L\equiv(u_L,d_L,s_L)^T$ and $\Psi_R\equiv(u_R,d_R,s_R)^T$, 
and $(a_L)^{\mu\alpha\beta}$ and $(a_R)^{\mu\alpha\beta}$ are matrices 
in $u,d,s$ flavor space of the form
\begin{align}
	\label{aLaRmatrices}
	(a_L)^{\mu\alpha\beta}\equiv
	&\left(
	\begin{array}{ccc}
		(a^{(5)}_V)_{11} & 0 & 0\\[0.5em]
		0 & (a^{(5)}_Q)_{11} & (a^{(5)}_Q)_{12}\\[0.5em]
		0 & (a^{(5)}_Q)_{12}^* & (a^{(5)}_Q)_{22}\\
	\end{array}
	\right)^{\mu\alpha\beta},\nonumber\\
	(a_R)^{\mu\alpha\beta}\equiv
	&\left(
	\begin{array}{ccc}
		(a^{(5)}_U)_{11} & 0 & 0\\[0.5em]
		0 & (a^{(5)}_D)_{11} & (a^{(5)}_D)_{12}\\[0.5em]
		0 & (a^{(5)}_D)_{12}^* & (a^{(5)}_D)_{22}\\
	\end{array}
	\right)^{\mu\alpha\beta}.
\end{align}
The zero entries in these matrices 
are a result of electric-charge conservation.
We have introduced the $(a^{(5)}_V)_{11}$ entry 
for generality, 
where the subscript $V$ has been arbitrarily selected 
to indicate independence from all other matrix entries.
But when the $a^{(5)}$ quark-level construction is restricted to $d=5$, 
as in Eq.~\rf{quarkSME}, 
contributions to $(a^{(5)}_V)_{11}-(a^{(5)}_Q)_{11}$ 
from Higgs expectation values are absent, 
forcing $(a^{(5)}_V)_{11}=(a^{(5)}_Q)_{11}$.  
%These two equal entries 
%on the diagonal of $(a_L)^{\mu\alpha\beta}$ 
%pertain to the original left-handed weak doublet.  

\subsection{Meson $\boldsymbol{d=5}$ Lagrangian from $\boldsymbol{\chi\rm PT}$}
%{\color{green}\subsection{Meson $\boldsymbol{d=5}$ Lagrangian from Chiral Symmetry Breaking}}

The key ideas behind ordinary $\chi$PT 
also cover Lagrangian contributions 
such as those in Eq.~\rf{udsLR}, 
and the general reasoning for their application 
in SME contexts has already been worked out~\cite{ChPT1,ChPT2}. 
The massless-quark Lagrangian's global SU(3)$_L\times$SU(3)$_R$ symmetry 
is spontaneously broken down to SU(3)$_V$.
The usual coset construction~\cite{XPT} then permits 
the representation of the corresponding Nambu-Goldstone bosons 
as the following $3\times 3$ matrix:
\beq
\label{Udef}
U(x)\equiv\exp\Bigg[i\dfrac{\lambda_a\phi_a(x)}{F_0}\Bigg]\,,
\eeq
where $F_0$ is the pseudoscalar decay constant in the SU(3)$_V$ chiral limit,
$\lambda_a$ with $a=1,\ldots, 8$ are the Gell-Mann matrices,
and the $\phi_a$ are identified with the observed pseudoscalar-meson octet
\beq
\label{Octet}
\lambda_a\phi_a=
\left(\begin{array}{ccc}
	\pi^{0}+\frac{1}{\sqrt{3}}\eta_8
	& 
	\sqrt{2}\pi^{+}
	&
	\sqrt{2}K^{+}%\quad
	\\[0.5em] %& \\
	%\quad 
	\sqrt{2}\pi^{-}
	& 
	-\pi^{0}+\frac{1}{\sqrt{3}}\eta_8
	&
	\sqrt{2}K^{0}%\quad
	\\[0.5em] %&\\
	%\quad 
	\sqrt {2}K^{-}
	&
	\sqrt{2}\bar{K}^0
	&
	-\frac{2}{\sqrt{3}}\eta_8
\end{array}
\right).
\eeq

With the chiral transformations 
implemented via 
$U(x)\to R U(x) L^{\dagger}$,
where $(L,R)\in$ SU(3)$_L\times$SU(3)$_R$, 
the idea is to build the lowest-order meson Lagrangian from $U$
such that it inherits the quark transformation properties  
under {\it C}, {\it P}, {\it T}, and SU(3)$_L\times$SU(3)$_R$.

The flavor-trace pieces 
\begin{align}\label{singlets}
	(a_L^1)^{\mu\alpha\beta}& \equiv \tfrac{1}{3}{\rm Tr}\big[(a_L)^{\mu\alpha\beta}\big]\,,\nonumber\\ 
	(a_R^1)^{\mu\alpha\beta}& \equiv \tfrac{1}{3}{\rm Tr}\big[(a_R)^{\mu\alpha\beta}\big]\,,
\end{align}
of the SME coefficient matrices $(a_L)^{\mu\alpha\beta}$ and $(a_R)^{\mu\alpha\beta}$ in Lagrangian~\rf{udsLR} 
transform as singlets under SU(3)$_L\times$SU(3)$_R$, 
and thus preserve chiral symmetry. 
The remaining flavor octets
\begin{align}\label{octets}
	(a_L^8)^{\mu\alpha\beta} & \equiv (a_L)^{\mu\alpha\beta}-(a_L^1)^{\mu\alpha\beta}\,,\nonumber\\ 
	(a_R^8)^{\mu\alpha\beta} & \equiv (a_R)^{\mu\alpha\beta}-(a_R^1)^{\mu\alpha\beta}\,,
\end{align}
violate chiral symmetry. 
But the Lagrangian~\rf{udsLR} remains invariant 
under a chiral transformation $(L,R)$ 
if accompanied by the replacements 
\begin{align}
	\label{octet_transf}
	(a_L^8)^{\mu\alpha\beta} & \to L(a_L^8)^{\mu\alpha\beta}L^{\dagger}\,,\nonumber\\
	(a_R^8)^{\mu\alpha\beta} & \to R(a_R^8)^{\mu\alpha\beta}R^{\dagger}\,,
\end{align}
a property that should also be reflected 
at the meson level.

The lowest-order Lagrangian contribution
that can be constructed from the fields $U(x)$, $U^{\dagger}(x)$,
the coefficients $(a_L^1)^{\mu\alpha\beta}$, $(a_R^1)^{\mu\alpha\beta}$, 
$(a_L^8)^{\mu\alpha\beta}$, $(a_R^8)^{\mu\alpha\beta}$, 
and derivatives with the above properties 
may be parametrized as:
\begin{align}\label{Lmeson1}
	\delta{\cal L}_{\phi} & = r_L^1 F_0{}^2 (a_L^1)^{\mu\alpha\beta} {\rm Tr}\big[i U^{\dagger}\partial_\mu\partial_\alpha\partial_\beta U+{\rm H.c.})\big]\nonumber\\
	& \phantom{={}} + r_R^1 F_0{}^2 (a_R^1)^{\mu\alpha\beta} {\rm Tr}\big[i U\partial_\mu\partial_\alpha\partial_\beta U^{\dagger}+{\rm H.c.}\big]\nonumber\\
	& \phantom{={}} + r_L^8 F_0{}^2  {\rm Tr}\big[i (a_L^8)^{\mu\alpha\beta} U^{\dagger}\partial_\mu\partial_\alpha\partial_\beta U+{\rm H.c.}\big]\nonumber\\
	& \phantom{={}} + r_R^8 F_0{}^2  {\rm Tr}\big[i (a_R^8)^{\mu\alpha\beta} U\partial_\mu\partial_\alpha\partial_\beta U^{\dagger}+{\rm H.c.}\big]\,.
\end{align}
Here, 
$r_L^1$, $r_R^1$, $r_L^8$, and $r_R^8$ 
are real-valued dimensionless constants, 
which are left undetermined in this approach,
but are expected to be of order unity. 
As our main focus concerns free-meson propagation,
we are primarily interested in the quadratic piece of $\delta{\cal L}_{\phi}$.
The first and the second line of this expression 
can then be dropped
because they amount to traces of commutators. 

To implement the additional constraints on $\delta{\cal L}_{\phi}$ 
that arise from {\it C}, {\it P}, and {\it T}, 
we decompose $(a_L^8)^{\mu\alpha\beta}$ and $(a_R^8)^{\mu\alpha\beta}$ 
as follows:
\begin{align}\label{cpt_Decomp}
	(a^8)^{\mu\alpha\beta}&\equiv\frac{1}{2}\big[(a_L^8)^{\mu\alpha\beta}+(a_R^8)^{\mu\alpha\beta}\big]\,,\nonumber\\	
	(b^8)^{\mu\alpha\beta}&\equiv\frac{1}{2}\big[(a_L^8)^{\mu\alpha\beta}-(a_R^8)^{\mu\alpha\beta}\big]\,.\nonumber\\
\end{align}
In Lagrangian~\rf{udsLR},
$(a^8)^{\mu\alpha\beta}$ and $(b^8)^{\mu\alpha\beta}$ 
parametrize quark operators with definite {\it C}, {\it P}, and {\it T} parities.
Those of $(a^8)^{\mu\alpha\beta}$ match Table~\ref{CPTproperties},
while $(b^8)^{\mu\alpha\beta}$ is associated 
with {\it C}-even operators 
and can therefore not contribute to our {\it C}-odd $\delta{\cal L}_{\phi}$. 
This forces the constraint $r_L^8=r_R^8\equiv\frac{1}{4}r_8$ 
on the couplings $r_L^8$ and $r_R^8$.
It is thus apparent 
that the leading-order CPT- and Lorentz-violating effects on the entire pseudoscalar-meson octet 
arising from underlying $a^{(5)}$ quark SME coefficients 
are predicted up to a single undetermined dimensionless parameter $r_8$.

To exhibit the resulting quadratic piece $\delta{\cal L}_{\phi}$ transparently,
we group the neutral- and charged-meson fields 
into column forms $\phi_{\rm N}\equiv(\pi^0,\eta,K^0)^T$ and $\phi_{\rm C}\equiv(K^+,\pi^+)^T$,
respectively.
This yields
\begin{align}\label{octet_L}
	\delta{\cal L}_{\phi} =
	& +i\frac{1}{2}r_8\, \phi_{\rm N}^{\dagger}\,(k^{(5)}_{\rm N})^{\mu\alpha\beta}\partial_\mu\partial_\alpha\partial_\beta\,\phi_{\rm N}+{\rm H.c.}\nonumber\\
	& +i\frac{1}{2}r_8\, \phi_{\rm C}^{\dagger}\,(k^{(5)}_{\rm C})^{\mu\alpha\beta}\partial_\mu\partial_\alpha\partial_\beta\,\phi_{\rm C}+{\rm H.c.},
\end{align}
where we have set
\begin{widetext}
	\begin{align}
		\label{neutral_scalars}
		(k^{(5)}_{\rm N})^{\mu\alpha\beta}=&
		\left(\!
		\begin{array}{ccc}
			0 & 0 & \frac{-1}{\sqrt{2}}\big[(a_D^{(5)}){}_{12}^*+(a_Q^{(5)}){}_{12}^*\big]  \\[0.5em]
			0 & 0 & \sqrt{\frac{3}{2}} \big[(a_D^{(5)}){}_{12}^*+(a_Q^{(5)}){}_{12}^*\big] \\[0.5em]
			\frac{-1}{\sqrt{2}}\big[(a_D^{(5)}){}_{12}+(a_Q^{(5)}){}_{12}\big]\; & 
			\;\sqrt{\frac{3}{2}} \big[(a_D^{(5)}){}_{12}+(a_Q^{(5)}){}_{12}\big]\; & \;
			(a_D^{(5)}){}_{22}-(a_D^{(5)}){}_{11}+(a_Q^{(5)}){}_{22}-(a_Q^{(5)}){}_{11} \\
		\end{array}
		\!\right)^{\mu\alpha\beta},\nonumber\\[1em]
		(k^{(5)}_{\rm C})^{\mu\alpha\beta}=&
		\left(\!
		\begin{array}{cc}
			(a_D^{(5)}){}_{22}+(a_Q^{(5)}){}_{22}-(a_U^{(5)}){}_{11}-(a_V^{(5)}){}_{11}\; & 
			(a_D^{(5)}){}_{12}+(a_Q^{(5)}){}_{12} \\
			(a_D^{(5)}){}_{12}^*+(a_Q^{(5)}){}_{12}^* & 
			\;(a_D^{(5)}){}_{11}+(a_Q^{(5)}){}_{11}-(a_U^{(5)}){}_{11}-(a_V^{(5)}){}_{11} \\
		\end{array}
		\!\right)^{\mu\alpha\beta}.
	\end{align}
\end{widetext}	
As required,
there is no mixing between electromagnetically neutral and charged mesons. 
Flavor-changing effects, 
however,
are included in $\delta{\cal L}_{\phi}$,
as evidenced by the nonzero off-diagonal elements 
contained in both $(k^{(5)}_{\rm N})^{\mu\alpha\beta}$ and $(k^{(5)}_{\rm C})^{\mu\alpha\beta}$. 
These are governed by the coefficients $(a_D^{(5)}){}_{12}$ and $(a_Q^{(5)}){}_{12}$,
which are associated with $d\leftrightarrow s$ mixing 
and may be thought of as a CPT- and Lorentz-violating contribution 
to the usual Cabibbo angle~\cite{ChPT2}. 
The $\pi^0$ and $\eta$ are composed of particle-antiparticle pairs in the quark model.
It is then intuitively reasonable 
that the spin-independent $a^{(5)}$ effects 
cancel at leading order between quark-antiquark pairs 
leading to the zero entries 
appearing in $(k^{(5)}_{\rm N})^{\mu\alpha\beta}$.
Setting $(a^{(5)}_V)_{11}=(a^{(5)}_Q)_{11}$ 
recovers the pure $d=5$ SME case.

The $(k^{(5)}_{\rm N})^{\mu\alpha\beta}$ coefficient reveals
that the SME corrections to the neutral-kaon system are twofold:
there is CPT- and Lorentz-breaking mixing with $\pi^0$ and $\eta$ mesons
as well as kinematic corrections confined to the kaon system itself. 
The former will affect the composition of the propagating kaon states 
as well as those of the $\pi^0$ and $\eta$ in a momentum-dependent way;
an investigation of the ensuing phenomenology 
and the associated potential for CPT and Lorentz tests
would be of interest,
but lies outside the scope of the present work. 
We will therefore proceed by setting  
$(a_D^{(5)}){}_{12}+(a_Q^{(5)}){}_{12}=0$ 
and focus entirely on the $(3,3)$ entry of $(k^{(5)}_{\rm N})^{\mu\alpha\beta}$. 
Its contribution to the kaon Lagrangian ${\cal L}_{K^0}$
is given by 
\beq\label{kaonL}
{\cal L}_{K^0}\supset\frac{1}{2}i\bar{K}^0(k^{(5)}_a)^{\mu\alpha\beta}_{K^0}\partial_\mu\partial_\alpha\partial_\beta K^0 +{\rm H.c.}\,,
\eeq
where
\beq\label{kdef}
(k^{(5)}_a)^{\mu\alpha\beta}_{K^0}=
r_8 \sum_{X=D,Q} \big[(a_X^{(5)}){}_{22}-(a_X^{(5)}){}_{11}\big]^{\mu\alpha\beta}.
\eeq
Note that this latter relation provides the desired explicit link  
between the $k^{(5)}$ introduced phenomenologically into pseudoscalar-meson field theory~\cite{ek19}
and $a^{(5)}$-type quark coefficients in the SME~\cite{Kostelecky:2018yfa}. 
In the next section,
we discuss phenomenological aspects of this correction 
and their implications for CPT and Lorentz tests 
involving neutral-meson oscillations.

To make contact with the full Dirac-spinor notation for the $a^{(5)}$ coefficient as in Eq.~(\ref{a5def}), 
we note that the $(d,d)$ and $(s,s)$ entries of $(a^8)^{\mu\alpha\beta}$ 
govern the behavior of the $d$ and $s$ quark, respectively,
and accordingly we denote them 
by $a_d^{(5)\mu\alpha\beta}$ and $a_s^{(5)\mu\alpha\beta}$. 
Equation~(\ref{kdef}) then takes the form
\beq\label{kdef2}
(k^{(5)}_a)^{\mu\alpha\beta}_{K^0}=
r_8 \big(a_s^{(5)\mu\alpha\beta}-a_d^{(5)\mu\alpha\beta}\big).
\eeq
Note that both $a_d^{(5)\mu\alpha\beta}$ and $a_s^{(5)\mu\alpha\beta}$ 
contribute with the same overall $r_8$ factor to this expression,
a feature that can be traced 
to our initial assumption of massless quarks
and the associated exact SU(3) flavor symmetry. 
The known nonzero quark masses will likely 
lead to corrections to this result of the form
\beq\label{kdef3}
(k^{(5)}_a)^{\mu\alpha\beta}_{K^0}=
r_8^s a_s^{(5)\mu\alpha\beta}-r_8^d a_d^{(5)\mu\alpha\beta}
\equiv\Delta a_{K^0}^{(5)\mu\alpha\beta},
\eeq
with the $d$ and $s$ quarks contributing differently. 
Paralleling the established notation for the $a^{(3)}$ case~(\ref{k3expr}),
we absorb the two $r$ factors into the definition of $\Delta a_{K^0}^{(5)}$. 
We recall that symmetry in all three indices yields 20 independent components.
The four tracelessness conditions $(k^{(5)}_a)^{\mu\alpha\beta}_{K^0}\eta_{\alpha\beta}=0$ 
reduce this number to 16 measurable coefficients at $d=5$~\cite{ek19}.

For mesons containing heavier quarks,
$\chi\rm PT$ methods become inapplicable,
as this situation no longer represents a perturbation 
around chiral symmetry.
However, 
based on the discrete symmetries 
associated with $a^{(5)}$
and its Lorentz-index structure,
$a^{(5)}$ effects in neutral mesons 
with heavier-quark content 
must presumably also be governed 
by relations analogous to Eq.~(\ref{kdef3}), 
albeit with larger differences between their respective two $r$ factors
and larger deviations from unity of these factors.
In this sense,
the general structure of Eq.~(\ref{kdef3})
is expected to remain valid 
for the $D^0$, $B^0_d$, and $B^0_s$ systems.

%
%However, 
%based on the discrete symmetries 
%associated with $a^{(5)}$
%and its Lorentz-index structure,
%we conjecture 
%that $a^{(5)}$ effects in neutral mesons 
%with heavier-quark content 
%are governed by relations analogous to Eq.~(\ref{kdef3}),
%albeit with larger differences between their respective two $r$ factors
%and larger deviations from unity of these factors.
	
%%%%%%%%%%%%%%%%%%%%%%%%%%%%%%%%%%%%%%%%%%%%%%%%%%%%%%%%%%%%%%%%%%%%%%%%%%%%%%%%%
	
\section{Phenomenology}
\label{nonmin}
	
As mentioned above,
neutral-meson oscillations 
provide experimental access to $\xi$. 
Thus, 
for the neutral $K^0$ meson 
consisting of the light $d$ and $s$ quarks, 
Eqs.~\rf{xi_d} and~\rf{kdef3} provide such access 
to various components of the $a^{(5)}_{K^0}$ microscopic SME coefficients:
\beq\label{xi5K0}
\xi^{(5)}_{K^0} = \frac{m^{2}}{\Delta\lambda}\zeta_{\rh}\zeta_{\si}\zeta_{\ta}\,\Delta a_{K^0}^{(5)\rho\sigma\tau}\,,
\eeq
with analogous relations expected to hold 
for the $D^0$, $B^0_d$, and $B^0_s$ systems.
We mention in passing 
that higher-order CPT violation would affect the eigenvalues of the $2 \times 2$ Hamiltonian~\rf{uvwx},
and thus meson oscillations. 
Although this is of theoretical significance, 
it is neglected in most studies 
due to the presumed small size of CPT breakdown. 
	
The mathematical structure of our SME expression for $\xi^{(5)}_{K^0}$ 
is analogous to that derived using a pseudoscalar-field theory~\cite{ek19},
as expected.
The right-hand side of Eq.~\rf{xi5K0} 
contains the kaon's 4-velocity $\zeta^{\mu}=\gamma(1,\vec{\beta})$, 
as measured relative to the coordinate system 
in which the components of the $a^{(5)}$ coefficients 
are specified. 
The $\xi^{(5)}_{K^0}$ on the left-hand side of Eq.~\rf{xi5K0} then represents the CPT- and Lorentz-violating correction
in the $2 \times 2$ effective Hamiltonian~\rf{uvwx}
that is comoving with the kaon in question.
We remark 
that, in principle, any inertial frame may serve 
as the $a^{(5)}$ coordinate system,
but the standard and practical choice 
is the sun-centered frame (SCF)~\cite{Mattcoords02}. 
The following four general features of $d=5$ CPT violation 
in the kaon system,
which are compatible with an underlying effective field theory, 
are then apparent. 
See also Table~\ref{gbfactors}.

First,
we observe that the directional dependence of $\zeta^{\mu}$ 
is determined by the kaon's four-momentum in the laboratory 
and the laboratory's orientation relative to the SCF. 
The laboratory's rotation about Earth's axis then implies 
that $\zeta^{\mu}$,
and thus the observable $\xi$,
will contain pieces 
with a time dependence 
governed by the sidereal frequency $\om_\oplus\simeq 2\pi(23\,\textrm{h}\;56\,\textrm{min})^{-1}$~\cite{NMPhKy99}. 
Since %up the% 
three symmetric factors of $\zeta^{\mu}$ 
appear in Eq.~\rf{xi5K0}, 
variations up to the third harmonic in $\omega$ 
are predicted. 
Time-stamped data 
are therefore highly desirable 
in experimental searches 
for this telltale signal of CPT violation.

A second aspect 
associated with the directional dependence of $\xi$ via $\zeta^{\mu}$ is 
that the various relevant laboratories differ in their orientation, 
they may collect data with differing kaon-momentum directions,
the kaon beam may or may not be collimated, 
etc. 
As a result, 
these experiments likely probe independent combinations of 
$\Delta a_{K^0}^{(5)\mu\alpha\beta}$ components 
and can correspondingly be viewed as independent CPT tests. 

As a third feature,
we note that in some experimental studies
two kaons
participating in a single process 
are compared,
so that two versions of the Hamiltonian~\rf{uvwx},
one for each kaon,
enter the theoretical description of such a system. This is the case, for instance, at the kaon factory at DA$\Phi$NE, where kaons are produced from $\Phi$ at resonance in a correlated state emerging in opposite directions. 
Since the kaons' four-velocities 
$(\zeta_1)^\mu\neq(\zeta_2)^\mu$
will in general differ from one another, 
predictions for observables 
may depend on two distinct CPT-violation parameters
$\xi_1\neq\xi_2$.
   
A fourth feature 
is that each of the three four-velocities appearing in Eq.~\rf{xi5K0} 
contains the usual relativistic boost factor $\gamma$.
The resulting $\gamma^3$ enhancement of the CPT-violation signal 
tracks well with the general effective-field-theory expectation 
that the relative importance of operators of higher mass dimension 
grows with momentum. 
Higher-energy kaons 
might therefore offer benefits for CPT tests~\cite{BestD015,KTeV,LHCb16}.
However, 
this advantage has to be weighed against 
better statistics 
and cleaner events
that lower-energy experiments 
might deliver~\cite{SchubBbar16,Babar08,KLOE,Belle12}.	
	
\begin{table}
	\begin{center}
		\tabcolsep8pt
		\begin{tabular}{ccc}
			\hline\hline
			Coefficient & Boost factors & Max.\ sidereal harmonic\\
			\hline
				
			$\De a^{(5)000}$ & $\ga^3$ & 0\\
				
			$\De a^{(5)00j}$ & $\ga^3\be_j$ & 1\\
				
			$\De a^{(5)0jk}$ & $\ga^3\be_j\be_k$ & 2\\
				
			$\De a^{(5)jkl}$ & $\ga^3\be_j\be_k\be_l$ & 3\\
			\hline\hline
		\end{tabular}
		\vskip -5pt
		\caption{Boost factors, directional dependence, and harmonics associated with the various components of $a^{(5)}$. 
				Here, 
				the Latin indices $j,k,l \in \{ X,Y,Z\}$ are spatial indices in the SCF. } 
		\label{gbfactors}
	\end{center}
	\vskip -10pt
\end{table}

To tailor our result~\rf{xi5K0} 
to particular experimental situations,
the meson's 4-velocity~\rf{beta_expr} 
needs to be specified.
This amounts to the determination of the kaon four-momentum $p^\mu$
because $p^\mu=m \zeta^{\mu}$
continues to hold at leading order in the SME. 

To write $\xi^{(5)}$ directly in terms of the customary SCF components of $a^{(5)}$,
the 4-velocity needs to be expressed in the same frame. 
We therefore continue 
by exhibiting the explicit coordinate changes 
required to transform between the laboratory and SCF systems. 
Due to the motion of Earth relative to the Solar System, 
terrestrial laboratories execute both rotations and boosts
relative to the SCF.
However, 
we may disregard the contributions from boosts 
since they are small compared to the size of rotational effects.
For this reason,
we are led to consider solely 
changes in the spatial orientations of laboratories 
as Earth rotates about its axis. 
	
\begin{figure}[t]
	\centering
		\includegraphics[width=1.0\linewidth]{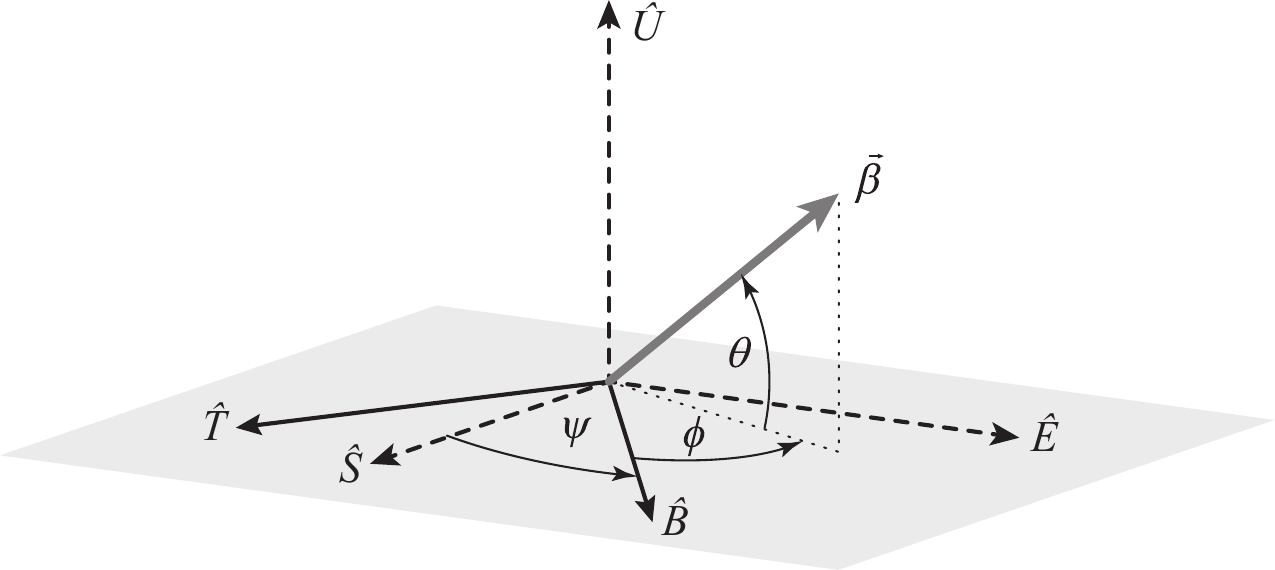}
		\caption{ 
			Conventions for setting up $\{\hat{T},\hat{B},\hat{U}\}$ coordinates.
			}
		\label{fig:chi}
\end{figure}
	
We define the laboratory coordinates 
with respect to the local south, east, and up directions 
characterized by the unit vectors $\hat{S}$, $\hat{E}$, and $\hat{U}$,
respectively.
In terms of the SCF basis $\{\hat{X},\hat{Y},\hat{Z}\}$,
the laboratory basis is
\bea
\hspace{-5mm}\hat{S} &=&
\hat{X}\cos\chi\cos\om_\oplus T_\oplus + \hat{Y}\cos\chi\sin\om_\oplus T_\oplus-\hat{Z}\sin\chi\,,
\nonumber\\
\hspace{-5mm}\hat{E} &=&
- \hat{X}\sin\om_\oplus T_\oplus+\hat{Y}\cos\om_\oplus T_\oplus\,,
\nonumber\\
\hspace{-5mm}\hat{U} &=&
\hat{X}\sin\chi\cos\om_\oplus T_\oplus + \hat{Y}\sin\chi\sin\om_\oplus T_\oplus+\hat{Z}\cos\chi\,.
\label{SEUZXY}
\eea 
In this transformation,
$\om_\oplus\simeq 2\pi(23\,\textrm{h}\;56\,\textrm{min})^{-1}$ denotes the sidereal angular frequency, 
and $T_\oplus$ is the local sidereal time. 
The transformation between the laboratory and SCF coordinates 
involves a constant tilt by the colatitude angle $\chi$ 
from the $Z$ axis of the SCF 
to the location of the ``up" direction at the respective laboratory~\cite{Ding:2016lwt}.
It becomes now straightforward
to express the 3-velocity $\vec{\beta}$ of the meson, 
and thus $\zeta^\mu$, 
in the laboratory coordinates $\{\hat{S},\hat{E},\hat{U}\}$. 

Although the $\{\hat{S},\hat{E},\hat{U}\}$ system is sufficient 
to extract SME meson-oscillation physics~\cite{ek19},
it can be more convenient in practical applications 
to use two angles $\phi,\theta$
to parametrize the direction of $\vec{\beta}$   
relative to the collider's beam direction $\hat{B}$. 
For beams propagating in a local horizontal plane,
we may then select another laboratory frame $\{\hat{T},\hat{B},\hat{U}\}$,
where $\hat{T}$ is horizontal and transverse to the beam direction $\hat{B}$, see Fig.~\ref{fig:chi}.
Measuring the beam direction $\psi$
counterclockwise relative to local south, 
\bea
\hat{T} &=&
\hat{S}\sin\psi - \hat{E}\cos\psi\,,\nonumber\\
\hat{B} &=&
\hat{S}\cos\psi+\hat{E}\sin\psi\,,
\label{BTUSEU}
\eea 
the basis $\{\hat{T},\hat{B},\hat{U}\}$ 
spans the desired right-handed laboratory system,
and we choose to parametrize the meson's 3-velocity direction
%{\color{red} (The expressions below are only correct because $\hat{v} \equiv \hat{\be}$.)}% 
$\hat{\beta}=\vec{\beta}/\beta$ as
\beq
\label{v_parametr}
\hat{\beta}=-\hat{T}\sin\phi \cos\theta+\hat{B}\cos\phi \cos\theta+\hat{U}\sin\theta\,.
\eeq
Note that with these conventions, 
we have $\phi\in[0,2\pi)$ and $\theta\in[-\half\pi,+\half\pi]$.
	
With Eqs.~\rf{SEUZXY} and~\rf{BTUSEU} at hand,
we can now write $\hat{\beta}$ in the SCF:
\beq\label{v_in_XYZ}
\hat{\beta} = G_X \hat{X}+G_Y\hat{Y}+c\,\hat{Z}\,.
\eeq
Here,
we have defined the orientation factors
\begin{align}\label{G_defs}
G_X ={}&a \cos\om_\oplus T_\oplus+b \sin\om_\oplus T_\oplus\,,\nonumber\\
G_Y ={}&a \sin\om_\oplus T_\oplus-b \cos\om_\oplus T_\oplus\,,
\end{align}
and the purely geometric constants 
\begin{align}\label{abc_defs}
a ={}&\cos \theta  \cos \psi  \cos \chi  \cos \phi -\cos \theta  \sin \psi  \cos \chi  \sin \phi  \nonumber\\
&+\sin \theta  \sin \chi\,,\nonumber\\
b ={}&-\cos \theta  \sin \psi  \cos \phi -\cos \theta  \cos \psi  \sin \phi \,,\nonumber\\
c ={}& \cos \theta  \sin \psi  \sin \chi  \sin \phi -\cos \theta  \cos \psi  \sin \chi  \cos \phi \nonumber\\
&+\sin \theta  \cos \chi \,.
\end{align}
With these results,
the 4-velocity~\rf{beta_expr},
which is needed to connect the observable~\rf{xi5K0} 
to the SME coefficient $\Delta a_{K^0}^{(5)\rho\sigma\tau}$,
can be constructed explicitly.
\flushcolsend

In addition to the two single-power sidereal-time dependencies
given in Eq.~\rf{G_defs},
the following higher products of $G_X$ and $G_Y$ 
and their corresponding sidereal-variation patterns
occur in $\zeta^{\rho}\zeta^{\sigma}\zeta^{\tau}$:
\begin{widetext}
\begin{align}
G_XG_X 
={}&
(a^2 + b^2) + (a^2 - b^2)\cos2\om_\oplus T_\oplus + 2ab\sin2\om_\oplus T_\oplus\,,
\nonumber\\
G_YG_Y 
={}&
(a^2 + b^2) + (a^2 - b^2)\cos2\om_\oplus T_\oplus - 2ab\sin2\om_\oplus T_\oplus\,,
\nonumber\\
G_XG_Y
={}&
ab\cos2\om_\oplus T_\oplus + \tfrac{1}{2}(a^2 - b^2)\sin2\om_\oplus T_\oplus\,,
\nonumber\\
G_XG_XG_Y
={}&
\tfrac{1}{4}\left[a(a^2-b^2)\sin\om_\oplus T_\oplus - b(a^2+b^2)\cos\om_\oplus T_\oplus\right] + \tfrac{1}{4}\left[a(a^2-3b^2)\sin3\om_\oplus T_\oplus + b(b^2-3a^2)\cos3\om_\oplus T_\oplus\right],
\nonumber\\
G_XG_YG_Y
={}&
\tfrac{1}{4}\left[b(a^2+b^2)\sin\om_\oplus T_\oplus + a(a^2+b^2)\cos\om_\oplus T_\oplus\right] + \tfrac{1}{4}\left[b(b^2-3a^2)\sin3\om_\oplus T_\oplus + a(3b^2-a^2)\cos3\om_\oplus T_\oplus\right],
\nonumber\\
G_XG_XG_X
={}&
\tfrac{3}{4}\left[b(b^2-a^2)\sin\om_\oplus T_\oplus + a(a^2+b^2)\cos\om_\oplus T_\oplus\right]+ \tfrac{1}{4}\left[b(3a^2-b^2)\sin3\om_\oplus T_\oplus + a(a^2-3b^2)\cos3\om_\oplus T_\oplus\right],
\nonumber\\
G_YG_YG_Y
={}&
\tfrac{3}{4}\left[a(a^2+b^2)\sin\om_\oplus T_\oplus - b(a^2+b^2)\cos\om_\oplus T_\oplus\right] + \tfrac{1}{4}\left[a(3b^2-a^2)\sin3\om_\oplus T_\oplus + b(3a^2-b^2)\cos3\om_\oplus T_\oplus\right].
\label{geomfuncs}
\end{align}
\end{widetext}
These expressions,
together with Table~\ref{orifactors5},
determine the sidereal dependence 
of each $\Delta a^{(5)}_{K^0}$ component. 
For example,
$\Delta a^{(5)}_{TXZ}$ 
contributes to $\xi^{(5)}_{K^0}$ via Eq.~\rf{xi5K0}
%in the summation over all indices in Eq.~\rf{xi5K0}
with the term $\frac{m^2}{\Delta\lambda}\gamma^3(c\, G_X)\Delta a^{(5)}_{TXZ}$,
where the geometric factors 
taken from Table~\ref{orifactors5} 
are shown in parentheses.
For completeness,
we display in Table~\ref{orifactors3} 
the corresponding results for $\Delta a^{(3)}$.

\begin{table}[H]
	\begin{center}
		\tabcolsep8pt
		\begin{tabular}{cc}
			\hline\hline
			\phantom{xxx} Coefficient \phantom{xxx} & \phantom{xxx} Orientation factors \phantom{xxx}\\
			\hline
			$\Delta a^{(5)}_{TTT}$ & 1\\
			$\Delta a^{(5)}_{TTZ}$ & c\\
			$\Delta a^{(5)}_{TZZ}$ & $c^2$\\
			$\Delta a^{(5)}_{ZZZ}$ & $c^3$\\
			$\Delta a^{(5)}_{TTX}$ & $G_X$\\
			$\Delta a^{(5)}_{TTY}$ & $G_Y$\\
			$\Delta a^{(5)}_{TXZ}$ & $c\, G_X$\\
			$\Delta a^{(5)}_{TYZ}$ & $c\, G_Y$\\
			$\Delta a^{(5)}_{XZZ}$ & $c^2\, G_X$\\
			$\Delta a^{(5)}_{YZZ}$ & $c^2\, G_Y$\\ 
			$\Delta a^{(5)}_{TXX}$ & $G_XG_X$\\
			$\Delta a^{(5)}_{TYY}$ & $G_YG_Y$\\
			$\Delta a^{(5)}_{XXZ}$ & $c\, G_XG_X$\\
			$\Delta a^{(5)}_{YYZ}$ & $c\, G_YG_Y$\\
			$\Delta a^{(5)}_{TXY}$ & $G_XG_Y$\\
			$\Delta a^{(5)}_{XYZ}$ & $c\, G_XG_Y$\\
			$\Delta a^{(5)}_{XXX}$ & $G_XG_XG_X$\\
			$\Delta a^{(5)}_{YYY}$ & $G_YG_YG_Y$\\
			$\Delta a^{(5)}_{XXY}$ & $G_XG_XG_Y$\\
			$\Delta a^{(5)}_{XYY}$ & $G_XG_YG_Y$\\
			\hline\hline
		\end{tabular}
		\vskip -5pt
		\caption{Orientation factors for $\Delta a^{(5)}$ components.}
		\label{orifactors5}
	\end{center}
	\vskip -10pt
\end{table}
	
Limits on $\Delta a^{(5)}$ 
may be obtained through further experimentation
or dedicated reanalyses of existing data. 
In what follows,
we list a few considerations pertaining to such experimental efforts.
We will specifically focus on KLOE,
but measurements at other facilities 
and with different neutral mesons
will require analogous conceptual analyses. 

At KLOE, 
kaons are produced nearly isotropically 
with a laboratory speed corresponding to $\be_{K_S} \approx 0.2$ with $\gamma \approx 1.03$~\cite{KLOEspecs}.
The orientation factors~\rf{geomfuncs}
depend on KLOE's colatitude $\chi=48.2^\circ$
and its beam direction $\psi=40^\circ$.  
The complete specification of these factors 
also requires  
the sidereal time $T_\oplus$
and the 3-velocity direction of the kaon $(\phi,\theta)$.
As these quantities are event specific, 
integration over all events  
would reduce coverage of the SME parameter space.
For example, 
time averaging 
precludes the observation of sidereal signals. 
Direction averaging 
likewise prevents detection of SME-induced time-dependent effects:
due to KLOE's isotropic momentum coverage 
different laboratory orientations associated with sidereal-time evolution
would become physically equivalent. 

These SME features suggest time and 3-direction binning 
of kaon measurements, 
as has indeed been undertaken 
in previous KLOE analyses of $\Delta a^{(3)}$
permitting constraints on the first sidereal harmonic.
However, 
since the present $\Delta a^{(5)}$ coefficient 
may induce sidereal variations up to the third harmonic,
optimization of time and orientation bins 
may be valuable. 
This optimization could potentially involve features beyond the number of bins:
orientation-bin choices exist 
in which the SCF solid-angle coverage of one bin 
evolves into the coverage of another bin
as Earth rotates.
To have truly independent bins,
they should be selected 
to minimize SCF bin overlap of this type.

\begin{table}[t]
	\begin{center}
		\tabcolsep8pt
		\begin{tabular}{cc}
			\hline\hline
			\phantom{xxx} Coefficient \phantom{xxx} & \phantom{xxx} Orientation factors \phantom{xxx}\\
			\hline
			$\Delta a^{(3)}_{T}$ & 1\\
			$\Delta a^{(3)}_{Z}$ & c\\
			$\Delta a^{(3)}_{X}$ & $G_X$\\
			$\Delta a^{(3)}_{Y}$ & $G_Y$\\
			\hline\hline
		\end{tabular}
		\vskip -5pt
		\caption{Orientation factors for $\Delta a^{(3)}$ components.}
		\label{orifactors3}
	\end{center}
	\vskip -20pt
\end{table}

KLOE produces meson pairs 
that consist of two back-to-back correlated kaons.
Care must be taken 
when analyzing the time evolution of such a system.
As mentioned above, 
the inherent momentum dependence of CPT violation~\rf{xi5K0}
deduced from the SME framework
mandates the introduction of an individual $\xi^{(5)}(\vec{p})$ 
for each of the two kaons~\cite{NMPhKy01}. 
Until the decay of the first kaon, 
the time evolution of this system
is given in terms of the difference and sum of these two directionally dependent $\xi^{(5)}(\vec{p})$. 

However, 
observation of the first kaon's decay 
yields information about it,
such as its momentum direction
or, in the case of a flavor-specific decay mode, 
its identity as $K^0$ or $\bar{K}^0$.
The initial correlation 
then implies 
that the other, remaining kaon state 
has reversed three-momentum direction, 
and hence fixes the corresponding $\xi^{(5)}$. 
Depending on the decay mode, 
it may also fix the surviving kaon's particle-antiparticle parity.
One possible analysis of observables
would therefore start at the first decay 
and focus on the remaining kaon. 
In particular, 
a feasible orientation analysis 
in a symmetric $\Phi$ factory 
should be conducted as a directional snapshot 
in conjunction with the decay product 
of each first-decaying particle of the correlated pair.
A probability analysis in the SME framework 
was instigated in Ref.~\cite{AR17}. 
Further exploration of this topic 
would be of definite interest,
but lies outside the scope of the present work.

%%%%%%%%%%%%%%%%%%%%%%%%%%%%%%%%%%%%%%%%%%%%%%%%%%%%%%%%%%%%%%%%%%%%%%%%%%%%%%%%%

\section{Bounds and Sensitivity Estimates}
\label{EstimKLOE}
    
This section is aimed at 
providing insight into the sensitivity levels 
that may be reached
with dedicated analyses of kaon interferometry.  
Previous experimental studies 
have limited their attention to sidereal variations 
up to the first harmonic,
and direction factors up to the first power of $\vec{\beta}$,
sufficient for their goal to investigate $\Delta a^{(3)}$. 
However,
$\Delta a^{(5)}$ predicts effects up to the third sidereal harmonic.
This suggests two different avenues 
for shedding light on attainable sensitivities.

The first avenue utilizes existing measurements 
conducted up to the first harmonic 
and up to the first power of $\vec{\beta}$.
These can be matched directly
to the component combinations 
$\{T,T,T\}$,
$\{T,T,Z\}$,
$\{T,T,X\}$,
and $\{T,T,Y\}$.
Due to their predicted constant effects,
the $\{T,T,T\}$ and $\{T,T,Z\}$ components are customarily treated separately. 
For a combination of the two, 
the best bound comes from collimated-kaon beam experiments~\cite{NMPhKy98}.
 
The leading limits on individual $T$ and $Z$ components 
were obtained at a $\Phi$ factory~\cite{KLOE}.
At the first harmonic, 
results are available from KTeV and KLOE for $\Delta a^{(3)}$ spatial components~\cite{KTeV,KLOE}, 
and these can be cross-related to our $\{T,T,X\}$ and $\{T,T,Y\}$ components of $\Delta a^{(5)}$.
These measurements have already been converted 
to the scalar $d=5$ coefficients $k_{a,K}^{(5)TTT}$, 
$k_{a,K}^{(5)TTZ}$,
$k_{a,K}^{(5)TTX}$,
and $k_{a,K}^{(5)TTY}$
in Ref.~\cite{ek19}.

Here,
we use our result~\rf{kdef3} 
to translate these results into constraints 
on the corresponding $\Delta a^{(5)}$ components.
For completeness 
we list these constraints 
as the first six entries in Table~\ref{estimtable}. 

The remaining $\Delta a^{(5)}$ components 
contribute to effects with higher sidereal harmonics, 
higher powers of $\vec{\beta}$,
or both.
Even though possible in principle, 
existing data have not yet been fitted
specifically to such time and direction dependencies, 
precluding direct translation to limits on those $\Delta a^{(5)}$ components. 
Projecting bounds on these coefficients 
therefore requires additional assumptions 
and should thus be interpreted 
as estimating sensitivities
attainable in future experimental analyses.

Let us begin by considering 
$\Delta a_K^{(5)TXX}$,
$\Delta a_K^{(5)TYY}$, 
and $\Delta a_K^{(5)TZZ}$, 
as well as 
$\Delta a_K^{(5)XXZ}$,
$\Delta a_K^{(5)YYZ}$, 
and $\Delta a_K^{(5)ZZZ}$. 
These enter the constant contribution to $\xi^{(5)}$ 
with ${\cal O}(1)$ geometric prefactors.
The first set of these comes with two powers of $\zeta^\mu$, 
and the second set with three powers. 
We proceed 
by taking $\langle \beta_j\rangle = {\cal O}(10^{-1})$ 
for each component of $\vec{\beta}$ 
averaged over a given spatial bin.
The details of the laboratory frame 
fixing the $j$ directions are irrelevant. 
Although this assumption is crude,
it can sensibly capture the basic orders of magnitude involved. 
First, 
take $\beta\coloneq|\vec{\beta}| \approx 0.2$ at KLOE,  
so the boost factor $\ga$ is close to unity,
and we may take $\ga^3\approx 1$.
Second,
spatial bins can be selected 
such that $\langle \beta_j\rangle \simeq \tfrac{1}{\sqrt{3}}\beta\simeq 10^{-1}$.
A sample bin choice of this kind 
could be achieved as follows:
take the sphere representing detector coverage,
and cut it with the $x$--$y$, $y$--$z$, and $z$--$x$ planes.
These cuts divide the surface of the sphere 
into eight equal-sized bins, 
each in the shape of an equilateral spherical triangle.

With numerical estimates for the sizes of the geometrical factors 
and the $\vec{\beta}$ components in place,
we can now match these six $\Delta a_K^{(5)}$ components
to a generic $\xi$ measurement. 
The Particle Data Group~\cite{ParticleDataGroup:2024cfk} 
lists experimental results for the kaon's $\xi$
in terms $\delta=\frac{1}{2}\xi$ 
with an accepted value from KTeV of~\cite{KTeV:2010sng}
\begin{align}
\re \de_{K} ={}& (2.5 \pm 2.3) \times 10^{-4}\,,\nonumber\\ 
\im \de_{K} ={}& (-1.5 \pm 1.6) \times 10^{-5}\,,
\label{pdgdelta}
\end{align} 
%{\color{blue} 
which represents a rectangle in the complex-$\xi$ plane.
%One can imagine in a plane of $\im \xi$ and $\re \xi$ a rectangular area given by the error ranges indicated about the measured values. 
In the SME framework,  
complex values of $\xi$ 
are restricted by~\cite{NMPhKy01}
\beq
\re{\xi}=-2\,\frac{\De m}{\De \ga}\,\im{\xi}\,,
%\re{\xi}=-(2\times \De m / \De \ga)\im{\xi},
\label{ReIm}
\eeq
where $\De m$ and $\De \ga$ are the mass and decay-rate differences of the two physical meson states. 
This condition represents a line 
through the origin of the complex-$\xi$ plane.
For our present purposes of estimating sensitivities,
we may therefore take SME predictions for $\xi$ 
to be experimentally constrained to lie 
on the section of the line 
that is located inside the rectangle~\rf{pdgdelta}.

The above considerations 
together with the observed values of $\De m$ and $\De \ga$~\cite{ParticleDataGroup:2024cfk} 
then imply 
that deviation $\de\xi$ from the CPT- and Lorentz-symmetric value $\xi=0$
can be at most
\beq
|\de \xi|< 8.8 \times 10^{-5}
\label{Dexiabs} 
\eeq
within the SME context. 
Combining this constraint
with the absolute value of Eq.~\rf{xi5K0}
\beq
|\xi| = \frac{m^2}{\sqrt{\De m^2 + \tfrac{1}{4}\De \ga^2}}\,\zeta_{\rho}\zeta_{\sigma}\zeta_{\tau}\, \De a^{(5)\rho\sigma\tau}_{K^0}\,,
\label{AbsxiSMEK} 
\eeq
the experimental values of $\De m$ and $\De \ga$, 
and our above considerations concerning $\langle \beta_j\rangle$
can then be used to obtain the desired sensitivity estimates.
%} 
%
%
%%Our leading-order SME formula~\rf{xi5K0} predicts a real $\xi$, so we need to match to the first line of Eq.~\rf{pdgdelta}.% 
%%Taking the maximally allowed deviation from zero,we may conservatively estimate $|\xi_{K^0}|\lsim 10^{-3}$.% 
%%Taken $|\im \xi| \lsim 10^{-5}$
%%Note that this value is compatible with KLOE measurements of $\re \de =(2.3 \pm 2.7) \times 10^{-4}$~\cite{DAmbrosio:2006hes}.%
%%Such a reasoning%  
%%yields a sensitivity level of {\color{green} $10^{-17}\,\textrm{GeV}^{-1}$} 
%There is a tracelessness constraint that comes from the fact that the traces at some particular mass dimension $d$ can be absorbed into a lower-dimension coefficient. Referring to detail discussion in \cite{ek19} for the kaon system this appears as interdependence of some of the coefficients. In this connection, those coefficients are handled differently.  
%%\color{red} Field redefinitions permit the traces of a coefficient at fixed d to be absorbed into coefficients at lower d [29], so all coefficients can be taken as traceless.}
For $\Delta a_K^{(5)TXX}$,
$\Delta a_K^{(5)TYY}$,
and	$\Delta a_K^{(5)TZZ}$
the sensitivity level is on the order of $10^{-16}\,\textrm{GeV}^{-1}$, 
while for 
$\Delta a_K^{(5)XXZ}$,
$\Delta a_K^{(5)YYZ}$, 
and $\Delta a_K^{(5)ZZZ}$ it is on the order of $10^{-15}$.
%Here, we have input 
%$m \simeq 0.5\ \mathrm{GeV}$ 
%and $\De\la \simeq (-3.5 \times 10^{-15}-\frac{i}{2}\times 7.1 \times 10^{-15})\ \mathrm{GeV}$
%as well as the KLOE values 
%$\ga=1.03$ and $\be=0.2$ 
%and rounded to the closest full order of magnitude. For the coefficient estimate, the absolute value of $\De \la$ has an overall $10^{-15}\ \mathrm{GeV}$ order of magnitude.

The remaining two sets of components 
$\Delta a_K^{(5)TXY}$,
$\Delta a_K^{(5)TYZ}$,
$\Delta a_K^{(5)TXZ}$,
and 
$\Delta a_K^{(5)XXX}$,
$\Delta a_K^{(5)YYY}$,
$\Delta a_K^{(5)XXY}$,
$\Delta a_K^{(5)XYY}$,
$\Delta a_K^{(5)XZZ}$,
$\Delta a_K^{(5)YZZ}$,
$\Delta a_K^{(5)XYZ}$,
come with two and three
powers of $\zeta^\mu$, 
respectively,
and only contribute to amplitudes of sinusoidal sidereal variations~\cite{ek19}.
To determine the sensitivity for those,
consider a time-integrated $\xi$ measurement. 
Although such averaging
precludes the direct observation of sidereal variations,
any putative oscillations about the measured mean value $\bar{\xi}$ of $\xi$
would likely enter the final result 
as a part of quoted statistical uncertainty $\Delta \xi$
as follows.
Suppose that, in addition to a true statistical uncertainty $\delta\xi$, 
the measured signal also contains a sinusoidal component with amplitude $\xi_0$ 
covering whole cycles of oscillation.
Most of the data would then fall into a band of size $2(\xi_0+\delta\xi)$ 
centered around the mean $\bar{\xi}$.
A statistical analysis of this band 
would then yield a perceived statistical uncertainty $\Delta \xi \simeq \xi_0+\delta\xi$ 
from which we may conclude
\beq\label{stat_unc}
\xi_0\lsim\Delta\xi\,.
\eeq
This suggests
that the amplitude of any sidereal harmonic 
is unlikely to be much larger 
than the perceived statistical error $\Delta\xi$ 
available in the literature,
%Based on the conservative assumption 
%that these amplitudes are constrained 
%by the maximally allowed deviation of $\xi$ from zero,
and crude $\Delta a^{(5)}$ sensitivity estimates 
may be determined. 
Paralleling the numerical calculation 
in the previous paragraph, 
we obtain a sensitivity level of $10^{-16}\,\textrm{GeV}^{-1}$ 
for $\Delta a_K^{(5)TXY}$,
$\Delta a_K^{(5)TXZ}$,
and	$\Delta a_K^{(5)TYZ}$,
and $10^{-15}\,\textrm{GeV}^{-1}$ 
for 
$\Delta a_K^{(5)XXX}$,
$\Delta a_K^{(5)YYY}$, 
$\Delta a_K^{(5)XXY}$,
$\Delta a_K^{(5)XYY}$, 
$\Delta a_K^{(5)XZZ}$,
$\Delta a_K^{(5)YZZ}$,
and $\Delta a_K^{(5)XYZ}$.
All sensitivity estimates
are collected in Table~\ref{estimtable}.
It is understood that these results are further constrained by the four tracelessness conditions 
$\Delta a^{(5)\mu\alpha\beta}_{K}\eta_{\alpha\beta}=0$. 

\renewcommand\arraystretch{1.4}
\begin{table}
	\setlength{\tabcolsep}{3pt}
	\begin{tabular}{llc}
		\hline
		\hline
		Coefficient & Sensitivity (GeV$^{-1}$) & Refs. \\
		\hline
		$	|\De a^{(5)TTT}_{K} - 0.97\De a^{(5)TTZ}_{K}|\phantom{xx}	$&$	< 2 \times 10^{-24}	$&	\cite{NMPhKy98}	\\	
		$	|\De a^{(5)TTX}_{K}|,~|\De a^{(5)TTY}_{K}|	$&$	< 1.8 \times 10^{-25}	$&	\cite{KTeV}	\\	
		$	\De a^{(5)TTT}_{K}	$&$	(-2.4 \pm 4.4) \times 10^{-17}	$&	\cite{KLOE}	\\	
		$	\De a^{(5)TTX}_{K}	$&$	(1.2 \pm 2.8) \times 10^{-18}	$&	\cite{KLOE}	\\	
		$	\De a^{(5)TTY}_{K}	$&$	(2.7 \pm 2.7) \times 10^{-18}	$&	\cite{KLOE}	\\	
		$	\De a^{(5)TTZ}_{K}	$&$	(4.2 \pm 3.0) \times 10^{-18}	$&	\cite{KLOE}	\\	
		$	\De a^{(5)TXX}_{K}	$&$ 10^{-16} $& \cite{DAmbrosio:2006hes} \\
		$   \De a^{(5)TYY}_{K}	$&$ 10^{-16} $& \cite{DAmbrosio:2006hes} \\
		$	\De a^{(5)TZZ}_{K}	$&$ 10^{-16} $& \cite{DAmbrosio:2006hes} \\
		$	\De a^{(5)TXY}_{K}	$&$ 10^{-16} $& \cite{DAmbrosio:2006hes} \\
		$   \De a^{(5)TXZ}_{K}	$&$ 10^{-16} $& \cite{DAmbrosio:2006hes} \\
		$   \De a^{(5)TYZ}_{K}	$&$ 10^{-16} $& \cite{DAmbrosio:2006hes} \\
		$   \De a^{(5)XXX}_{K}	$&$ 10^{-15} $& \cite{DAmbrosio:2006hes} \\
		$   \De a^{(5)YYY}_{K}	$&$ 10^{-15} $& \cite{DAmbrosio:2006hes} \\
		$   \De a^{(5)ZZZ}_{K}	$&$ 10^{-15} $& \cite{DAmbrosio:2006hes} \\
		$   \De a^{(5)XXY}_{K}	$&$ 10^{-15} $& \cite{DAmbrosio:2006hes} \\
		$   \De a^{(5)XXZ}_{K}	$&$ 10^{-15} $& \cite{DAmbrosio:2006hes} \\
		$   \De a^{(5)XYY}_{K}	$&$ 10^{-15} $& \cite{DAmbrosio:2006hes} \\
		$   \De a^{(5)YYZ}_{K}	$&$ 10^{-15} $& \cite{DAmbrosio:2006hes} \\
		$   \De a^{(5)XZZ}_{K}	$&$ 10^{-15} $& \cite{DAmbrosio:2006hes} \\
		$   \De a^{(5)YZZ}_{K}	$&$ 10^{-15} $& \cite{DAmbrosio:2006hes} \\
		$   \De a^{(5)XYZ}_{K}	$&$ 10^{-15} $& \cite{DAmbrosio:2006hes} \\
		\hline
		\hline
	\end{tabular}
	\caption{Bounds and sensitivity estimates for $\De a_{K}^{(5)\mu\nu\rho}$.
	The first six rows contain two-sided bounds on $\De a_{K}^{(5)\mu\nu\rho}$ components 
	translated directly from Ref.\ \cite{ek19} using Eq.~(\ref{kdef3}). 
	The remaining rows show two-sided sensitivities 
	attainable via reanalysis of existing data.}
	\label{estimtable}
\end{table}

%%%%%%%%%%%%%%%%%%%%%%%%%%%%%%%%%%%%%%%%%%%%%%%%%%%%%%%%%%%%%%%%%%%%%%%%%%%%%%%%% 
 
\section{Conclusions}
\label{Conclusion}
	
In this work, 
we have investigated quark-level CPT and Lorentz violation in neutral-meson interferometry 
within the framework of the nonminimal SME. 
Starting from the generic mass-dimension five fermion operator 
involving the coefficient $a^{(5){\mu\alpha\beta}}$, 
we embedded this structure into the SME's quark sector 
while preserving gauge invariance 
and employed an adaptation of chiral perturbation theory 
to derive the corresponding leading-order contribution to the pseudoscalar-meson effective Lagrangian. 
This analysis established a concrete map 
between the microscopic SME quark coefficients 
and the effective scalar-field coefficients $k^{(5)\mu\alpha\beta}_a$ 
used in an existing phenomenological study of neutral-meson oscillations, 
complementing previously used valence-quark considerations. 
In particular, 
we obtained an explicit expression for the effective kaon coefficient $k_a^{(5){\mu\alpha\beta}}$ 
in terms of the down- and strange-quark coefficients.
This result clarifies the number and type of independent observable combinations at $d=5$
and determines directly the CPT-violating parameter $\xi^{(5)}$. 
Although various details of this reasoning 
become inapplicable for heavier quarks,
general symmetry considerations indicate analogous results 
for the $D^0$, $B^0_d$, and $B^0_s$ meson systems.
	
On the phenomenological side, 
we have shown that the resulting contribution~\rf{xi5K0}
inherits the full orientation and boost dependence 
required by realistic CPT-violating effective field theory, 
implying sidereal variations up to the third harmonic 
and a characteristic $\gamma^3$ enhancement of CPT-violating signals. 
We have illustrated these general features 
by expressing observables in the SCF, 
deriving explicit orientation factors 
tailored to realistic laboratory configurations and, 
as a specific example, 
to the geometry of the KLOE detector. 
This allows a systematic identification of the combinations of $\Delta a^{(5)}_{K}$ 
that are probed by different beam orientations, 
laboratory locations, 
and momentum bins 
with time-stamped data. 
This result highlights 
how experiments with distinct geometries or beam configurations 
can access complementary regions of the SME coefficient space.
	
Finally, 
we have discussed the implications 
for the example of correlated-kaon measurements 
in a $\Phi$-factory environment 
and provided order-of-magnitude sensitivity estimates 
suggesting that dedicated measurements or reanalyses of existing data, 
optimized for detecting higher sidereal harmonics, 
would likely reach sensitivities of $10^{-15\ldots-16}\,\text{GeV}^{-1}$ or better 
to mass dimension $d=5$ CPT- and Lorentz-violating quark coefficients in the SME. 
Our results indicate that neutral-meson interferometry 
remains a powerful test of CPT and Lorentz symmetry 
beyond the minimal SME, 
furnishing a bridge between quark-level SME coefficients 
and experimentally accessible observables 
that can be extended to heavier neutral-meson systems 
and to higher-dimensional operators in future work.

%%%%%%%%%%%%%%%%%%%%%%%%%%%%%%%%%%%%%%%%%%%%%%%%%%%%%%%%%%%%%%%%%%%%%%%%%%%%%%%%%

\end{document}